\documentclass[twocolumn,twocolappendix,usenatbib,useAMS,fleqn,usenames,dvipsnames,tighten]{aastex63}
\usepackage{amsmath}
\hypersetup{linkcolor=MidnightBlue,citecolor=MidnightBlue,filecolor=MidnightBlue,urlcolor=MidnightBlue}
\graphicspath{{./}{figures/}{fig_m12f/}{fig_m12i/}}
\accepted{}
\submitjournal{MNRAS}
\usepackage{enumitem,amsmath,amstext,mathtools}

\usepackage{cancel}

\newcommand{\datastatement}[1]{\begin{small}\section*{Data Availability Statement}\end{small}{\noindent #1}\vspace{0pt}}

\newcommand{\be}{\begin{equation}}
\newcommand{\ee}{\end{equation}}

\newcommand{\lstar}{L_{\ast}}

\newcommand{\Alf}{{Alfv\'en}}
\newcommand{\kmin}{{\kappa_{\rm eff}^{\rm min}}}

\begin{document}

\shorttitle{Constraining CR Transport in the CGM}
\shortauthors{Butsky et al. }

\title{Constraining Cosmic-ray Transport with Observations of the Circumgalactic Medium}
\correspondingauthor{Iryna S. Butsky}
\email{ibutsky@caltech.edu}

\author[0000-0003-1257-5007]{Iryna S. Butsky}
\affiliation{TAPIR, California Institute of Technology, Pasadena, CA 91125, USA}
\author{Shreya Nakum}
\affiliation{Milpitas High School, Milpitas, CA 95035}
\author[0000-0002-7484-2695]{Sam B. Ponnada}
\affiliation{TAPIR, California Institute of Technology, Pasadena, CA 91125, USA}
\author[0000-0002-3817-8133]{Cameron B. Hummels}
\affiliation{TAPIR, California Institute of Technology, Pasadena, CA 91125, USA}
\author[0000-0001-9658-0588]{Suoqing Ji}
\affiliation{Shanghai Astronomical Observatory, Chinese Academy of Sciences, Shanghai 200030, China}
\author[0000-0003-3729-1684]{Philip F. Hopkins}
\affiliation{TAPIR, California Institute of Technology, Pasadena, CA 91125, USA}


\begin{abstract}
Recent theoretical studies predict that the circumgalactic medium (CGM) around low-redshift, $\sim \lstar$ galaxies could have substantial nonthermal pressure support in the form of cosmic rays. However, these predictions are sensitive to the specific model of cosmic-ray transport employed, which is theoretically and observationally underconstrained. In this work, we propose a novel observational constraint for calculating the lower limit of the radially-averaged, effective cosmic-ray transport rate, $\kmin$. Under a wide range of assumptions (so long as cosmic rays do not lose a significant fraction of their energy in the galactic disk, regardless of whether the cosmic-ray pressure is important or not in the CGM), we demonstrate a well-defined relationship between $\kmin$ and three observable galaxy properties: the total hydrogen column density, the average star formation rate, and the gas circular velocity. We use a suite of FIRE-2 galaxy simulations with a variety of cosmic-ray transport physics to demonstrate that our analytic model of $\kmin$ is a robust lower limit of the true cosmic-ray transport rate. We then apply our new model to calculate $\kmin$ for galaxies in the COS-Halos sample, and confirm this already reveals strong evidence for an effective transport rate which rises rapidly away from the interstellar medium to values $\kmin \gtrsim 10^{30-31}\,{\rm cm}^2\,{\rm s}^{-1}$ (corresponding to anisotropic streaming velocities of $v^{\rm stream}_{\rm eff} \gtrsim 1000\,{\rm km}\,{\rm s}^{-1}$) in the diffuse CGM, at impact parameters larger than $50-100$\,kpc. We discuss how future observations can provide qualitatively new constraints in our understanding of cosmic rays in the CGM and intergalactic medium.
\end{abstract}

\section{Introduction}
\label{sec:intro}

Cosmic rays have the potential to fundamentally alter the structure of the circumgalactic medium (CGM). The build-up of supernova-injected cosmic-ray pressure in the galactic disk drives mass-loaded galactic outflows \citep[e.g.,][]{Uhlig:2012, Booth:2013, Ruszkowski:2017, Wiener:2017, Girichidis:2018, Farber:2018, Hopkins:2021_outflows} that fill the CGM with metal-rich gas. Once in the CGM, cosmic rays provide nonthermal pressure support that affects the ionization state \citep{Salem:2016, Butsky:2018}, density \citep{Butsky:2020, Ji:2020}, and kinematics \citep{Buck:2020, Ji:2021, Butsky:2022} of multiphase CGM gas. These phenomena are most pronounced in the CGM around low-redshift, $\lstar$ galaxies, in which recent numerical studies show that cosmic-ray pressure may even exceed thermal pressure by orders of magnitude \citep{Salem:2016, Butsky:2018, Buck:2020, Ji:2020, Butsky:2022}. 

The predictive power of existing hydrodynamic simulations that include cosmic rays is hindered by the tremendous uncertainties in modeling cosmic-ray transport. Simulations with relatively small differences in cosmic-ray transport models (or even simulations modeling the same cosmic-ray transport with different numerical implementations) can produce a wide variety of galaxy properties \citep[e.g.,][]{Buck:2020, Gupta:2021, Hopkins:2021_transport2, Semenov:2022}. These differences are amplified in the CGM \citep{Butsky:2018}, even amongst cosmic-ray transport models that are consistent with observational constraints in the galactic disk \citep{Hopkins:2021_transport1}.

The main difficulty in constraining the effects of cosmic rays in the CGM around $\lstar$ galaxies is the lack of observational constraints. Almost all of the ``classic'' constraints on cosmic rays from Solar System observations (secondary-to-primary ratios, radioactive isotopes, matter versus antimatter, etc.), as well as indirect constraints, such as spatially-resolved galactic and extra-galactic $\gamma$-ray emission, or cosmic-ray ionization in nearby molecular clouds, constrain cosmic-ray propagation only in the typical interstellar medium (ISM; largely in Solar-neighborhood-type conditions), and contain almost no information about the CGM \citep[see e.g.][and references therein]{Chan:2019,Bustard:2020,delaTorre:2021.dragon2.methods.new.model.comparison,hopkins:cr.multibin.mw.comparison, Chan:2022,hopkins:2021.sc.et.models.incompatible.obs}. Synchrotron data, while interesting in its own right, constrains only the electron and positron populations, which contain a negligible fraction of the cosmic-ray pressure (which is almost entirely in $\sim\,1-10\,$GeV protons) and have qualitatively different loss processes and rates. While gamma-ray observations hint at the presence of cosmic-ray protons in the CGM around the Milky Way and M31 \citep[e.g.,][]{Feldmann:2013, Jana:2020, Do:2021, Recchia:2021}, existing observations do not sufficiently constrain models of cosmic-ray transport. 
This means that models where cosmic rays are trapped in the CGM and build up a cosmic-ray-pressure-supported halo, or models where cosmic rays effectively ``decouple'' and rapidly escape once out of the ISM, are at present equally allowed.

In this work, we propose a new approach for using observations of the CGM to place a lower limit on the effective cosmic-ray transport rate, $\kmin$. We consider a simple model of a spherical CGM in which supernovae inject cosmic rays at a rate proportional to the galaxy's star formation rate. The cosmic rays then move away from their injection site, out into the CGM, with a radially-averaged effective transport rate, $\kappa_{\rm eff}$. The value of $\kappa_{\rm eff}$ sets the shape of the cosmic-ray pressure gradient, which then determines how effectively cosmic rays can support gas against gravity. As we will show, the lower limit of the effective transport rate, $\kmin$, can be determined by three parameters that can feasibly be determined from observations: the total hydrogen column density, the star formation rate, and the circular velocity.

This rest of this work is organized as follows. In \autoref{sec:model}, we present the analytic model and discuss the regimes in which it is valid. In \autoref{sec:sims}, we demonstrate the success of this model in cosmological simulations and apply it to observations in \autoref{sec:obs}. We discuss our conclusions in \autoref{sec:conclusions}.

\section{Derivation \&\ Model Tests}
\label{sec:model}

\subsection{Assumptions \&\ Key Scalings}
\label{sec:model:analytic}
Consider a spherically-symmetric CGM around a galaxy with galactocentric radii, $r$. Supernovae inject cosmic-ray energy, $E_{\rm cr}$, into the ISM at a rate proportional to the star formation rate, $\dot{M}_{\ast}$, $\dot{E}_{\rm cr} = \epsilon_{\rm cr}\dot{E}_{\rm SNe} = \epsilon_{\rm cr}\bar{u}_{\rm SNe}\dot{M}_{\ast}$, where $\epsilon_{\rm cr} \sim 0.1$ \citep{Ginzburg:1964, TerHaar:1950, Gabici:2019} is the fraction of the total supernova energy, $E_{\rm SNe}$, that goes into cosmic-ray energy, and $\bar{u}_{\rm SNe} \sim 10^{51} {\rm erg}/100 M_{\odot}$ is the average supernova energy. 

Once formed, the cosmic rays propagate around tangled magnetic field lines away from their injection site. Assuming that on large scales, magnetic fields are either sufficiently tangled or symmetric, we can parameterize cosmic-ray transport in the outward radial direction assuming an effective radially-averaged cosmic-ray transport rate, $\kappa_{\rm eff}$, 
\begin{equation}
   \kappa_{\rm eff}(r) \equiv \langle |{\bf F}_{\rm cr}\cdot \hat{r}| / |\nabla e_{\rm cr}| \rangle,
\end{equation}
where ${\bf F}_{\rm cr}$ is the cosmic-ray flux and $e_{\rm cr}$ is the cosmic-ray energy density. While referring to the effective cosmic-ray transport rate as $\kappa_{\rm eff}$ is reminiscent of conventions for describing cosmic-ray diffusion, this assumption does not require that cosmic-ray transport actually obey a constant-$\kappa$ diffusion equation. For example, if cosmic-ray motion is advective with velocity $v_{\rm cr}$, then $\kappa_{\rm eff} \sim v_{\rm cr}e_{\rm cr}/|\nabla e_{\rm cr}|$. Simply put, $\kappa_{\rm eff}$ captures the effective isotropic cosmic-ray transport rate, while being agnostic about the detailed microphysics that determine that transport rate. 

We assume that cosmic rays propagate sufficiently fast to escape the ISM with negligible collisional or collisionless losses (as required by observations for dwarf and $\sim \lstar$ galaxies; see \citealt{lacki:2011.cosmic.ray.sub.calorimetric,lopez:2018.smc.below.calorimetric.crs,Chan:2019,Hopkins:2020}). Therefore, $\dot{E}_{\rm cr}$ remains a reasonable estimate of the total cosmic-ray energy injection rate. Finally, we assume that cosmic rays behave as an ultra-relativistic fluid with pressure $P_{\rm cr} \approx (\gamma_{\rm cr} - 1) e_{\rm cr}$, where $\gamma_{\rm cr} = 4/3$ is the cosmic-ray adiabatic index.

At large distances from the galactic disk ($r \gg 10\,$kpc), the dynamical and cosmic-ray propagation timescales are sufficiently long that small timescale variations ($\lesssim$ Gyr) in $\dot{E}_{\rm cr}$ are averaged out. Therefore, we approximate the galaxy as a point-like source of cosmic-ray energy, from which cosmic rays escape, and create an outward pressure gradient force, $\partial P_{\rm cr} / \partial r$, that counteracts gravity and contributes to maintaining hydrostatic equilibrium.\footnote{As long as $\kappa_{\rm eff} \lesssim 10^{35}\, {\rm cm}^2\,{\rm s}^{-1}$, the cosmic-ray ``scattering'' length is smaller than the halo, so cosmic rays exert pressure.} In steady-state, for any spherical shell around the galaxy at radius $r$, the flux of cosmic rays injected into the system by supernovae is balanced by the flux of cosmic rays ``diffusing'' out, and the global cosmic-ray pressure gradient remains unchanged, 
\begin{equation}
\label{eqn:pcr.steady.state}
    \dot{E}_{\rm cr} = \int_{\rm V} \nabla \cdot {\bf F}_{\rm cr} = 4 \pi r^2 \kappa_{\rm eff} \nabla e_{\rm cr}.
\end{equation}
Rearranging the terms in the above equation, and making the assumption that $\nabla e_{\rm cr} \approx 3 P_{\rm cr} / r$, we can determine the cosmic-ray pressure profile, 
\begin{equation}
    \label{eqn:pcr.profile}
    P_{\rm cr} = \frac{\dot{E}_{\rm cr}}{12 \pi \kappa_{\rm eff} r}. 
\end{equation}

Assuming the system is in quasi-static equilibrium, the inward gravitational force, $\rho_{\rm gas}\, \partial \Phi/\partial r = \rho_{\rm gas} V_{\rm c}^2/r$, should be balanced by the total outward pressure force, $\partial P_{\rm tot}/\partial r$. We define $\rho_{\rm gas}$ to be the gas density, $\partial \Phi/\partial r$ to be the gravitational acceleration, and $V_{\rm c}$ to be the average circular velocity at some galactocentric radius, $r$. The total pressure, $P_{\rm tot}$, encompasses the contributions of thermal, cosmic ray, magnetic, and turbulent pressures. 

First, we assume that cosmic-ray pressure is the dominant source of CGM pressure, $P_{\rm tot} \approx P_{\rm cr}$, and derive an expression for $\kappa_{\rm eff}$ in this limit.  Later, we will show that lifting this assumption means that the derived $\kappa_{\rm eff}$ is really a lower limit of the true value. 

Setting the gravitational force equal to the cosmic-ray pressure gradient and rearranging \autoref{eqn:pcr.profile}, we define a critical gas density at which gas is in quasi-static equilibrium,

\begin{equation}
\label{eqn:rho} \rho_{\rm crit}(r) \approx \frac{\dot{E}_{\rm cr}}{12\pi\, r\,V_{\rm c}^{2}\,\kappa_{\rm eff}}. 
\end{equation}

As discussed in \citet{Ji:2020} and \citet{Hopkins:2021_outflows}, this critical gas density profile is an equilibrium solution in the limit where cosmic rays dominate the total pressure.  If $\rho_{\rm gas} > \rho_{\rm crit}$, then $\rho_{\rm gas}\,|\partial \Phi/\partial r| > | \partial P_{\rm cr}/\partial r |$, so gravity would win and the gas would sink and accrete onto the galaxy. Conversely, if $\rho_{\rm gas} < \rho_{\rm crit}$, cosmic-ray pressure would accelerate gas outwards. However, since cosmic rays are diffusive, they would not ``sink'' with the gas. Because the gravitational potential is dominated by dark matter at large radii, and the cosmic-ray ``injection rate'', $\dot{E}_{\rm cr}$, is smoothed over the diffusion timescale to large (CGM) radii, the cosmic-ray profile should remain in steady-state even if accretion rates or galaxy dynamics vary. This means that the CGM gas density should relatively quickly reach an equilibrium value, with $\rho \approx \rho_{\rm crit}$, independent of the thermodynamic or other properties of the gas in the halo.

It is useful to express the critical gas density in terms of the equivalent projected surface density, $\Sigma_{\rm crit} \equiv \int \rho_{\rm crit}\,dl$ (or nucleon column density $N_{\rm H,\,crit}$). Integrating both sides of \autoref{eqn:rho} along a sightline $l$ that intersects the CGM at impact parameter $R$, we find
\begin{equation}
\label{eqn:column}
\Sigma_{\rm crit} \equiv m_{\rm p}\,N_{\rm H,\,crit} \approx \frac{f\,\dot{E}_{\rm cr}}{12\pi V_{\rm c}^{2}(R)\,\kappa_{\rm eff}(R)},
\end{equation}
where $f \approx 1$ is a dimensionless integral which varies weakly over the physical parameter space of interest.
Assuming $\dot{E}_{\rm cr}$ is dominated by supernovae, rearranging \autoref{eqn:column} gives us an estimate for the effective cosmic-ray transport rate at impact parameter $R$,
\begin{equation}
\label{eqn:kappa.vs.NH} \frac{\kappa_{\rm eff}(R)}{10^{30}\,{\rm cm^{2}\,s^{-1}}} \sim \frac{10^{18}{\rm cm^{-2}}}{N_{\rm H,\,crit}}\left[ \frac{\dot{M}_{\ast}}{\rm M_{\odot}\,yr^{-1}} \right]\left[ \frac{200\,{\rm km\,s^{-1}}}{V_{\rm c}(R)}\right]^{2}.
\end{equation}
In this case, $\kappa_{\rm eff}$ describes the total effective cosmic-ray transport rate, which includes advective motion with the gas as well as cosmic-ray motion relative to the gas. 

In a cosmic-ray-pressure-supported halo, where $P_{\rm tot} \approx P_{\rm cr}$, we expect $N_{\rm H} \approx N_{\rm H,\,crit}$ (up to a small normalization correction). However, if cosmic rays are sub-dominant to other forms of pressure (e.g., gas thermal pressure, free-fall or rotation, turbulent motions, magnetic pressure), then hydrostatic equilibrium implies that the halo can support higher gas densities against gravity, meaning that we will generally have $N_{\rm H} \gtrsim N_{\rm H,\,crit}$. This means the $\kappa_{\rm eff}$ inferred assuming $N_{\rm H} \sim N_{\rm H,\,crit}$ is really a {\em lower limit} to the true $\kappa_{\rm eff}$.  Therefore, this suggests that a robust {\em lower limit} to $\kappa_{\rm eff}$ can be estimated as $\kappa_{\rm eff} \ge \kmin = \kappa_{\rm eff}(N_{\rm H}=N_{\rm H,\,crit})$

\begin{equation}
\label{eqn:kappa.min} \frac{\kmin (R)}{10^{30}\,{\rm cm^{2}\,s^{-1}}} \sim  \frac{10^{18}\,{\rm cm^{-2}}}{N_{\rm H}} \,\left[ \frac{\dot{M}_{\ast}}{\rm M_{\odot}\,yr^{-1}} \right]\,\left[ \frac{200\,{\rm km\,s^{-1}}}{V_{\rm c}(R)}\right]^{2}.
\end{equation}

Summarized simply, if the value of $\kappa_{\rm eff}$ were smaller than $\kappa_{\rm eff}^{\rm min}$, then the observed column of gas $N_{\rm H}$ would be accelerated outwards and ejected by cosmic-ray pressure in approximately a free-fall time.

\subsection{Regimes of Validity}
\label{sec:model:valid}

Most of the assumptions in deriving \autoref{eqn:kappa.min} introduce only order-unity uncertainties that are small compared to the uncertainties in physics that set the cosmic-ray transport rate, or observational uncertainties in $N_{\rm H}$. This includes, for example, uncertainties in the spectral shape of the cosmic rays, small-scale fluctuations in $\kappa_{\rm eff}$, anisotropic transport along globally-ordered magnetic field lines, uncertainties in the true cosmic-ray injection fraction or supernova rate per unit stellar mass, details of the shape of the gravitational potential, and deviations from spherical symmetry. 

In most cases, breaking the assumptions we have made would result in the measured $\kmin$ being more of a lower limit. For example, as discussed above, our estimate of $\kmin$ as a {\em lower limit} to the true cosmic-ray $\kappa_{\rm eff}$ remains valid {\em even if the galaxies have negligible cosmic-ray pressure in the CGM}. Consider that, in order to (approximately) maintain hydrostatic equilibrium ($P_{\rm tot} \approx P_{\rm grav}$) the total pressure profile is set by the shape of the gravitational potential and remains the same regardless of which forms of pressure are dominant. Therefore, if cosmic-ray pressure in the CGM is not dominant (or even negligible), $P_{\rm cr} < P_{\rm tot}$, then given the inverse scaling of $P_{\rm cr}$ and $\kappa_{\rm eff}$ (\autoref{eqn:pcr.profile}), the observationally-measured $\kmin$ would be lower than the true $\kappa_{\rm eff}$, and \autoref{eqn:kappa.min} would still hold as a lower limit. We have validated this directly in simulations in which cosmic-ray pressure is a negligible fraction of the total CGM pressure, discussed below.

Another such assumption is that cosmic-rays are only accelerated by galactic supernovae. If galaxies host active galactic nuclei (AGN) with kinetic luminosities much larger than the supernova rate (for example, in galaxies with very massive bulges and very low star formation rates), then the cosmic-ray injection rate can be dominated by AGN, and no longer be proportional to the star formation rate. In this case, the ``true'' cosmic-ray injection rate, $\dot{E}_{\rm cr}$, would actually be {\em larger} than our estimate based on $\dot{M}_{\ast}$ (because of the AGN contribution) and our $\kappa_{\rm eff}^{\rm min}$ would again remain perfectly valid as a lower limit to the true $\kappa_{\rm eff}$. For additional examples, we refer the reader to \citet{Hopkins:2020, Hopkins:2022_subgrid}.

Recall that for a given cosmic-ray injection rate, a lower $\kappa_{\rm eff}$ at a given $r$ implies a higher $P_{\rm cr}$. Therefore, what we are effectively constraining is an upper limit to the total pressure, $P_{\rm tot}$, since exceeding some threshold value would blow out all the observed gas on the order of a freefall time. In principle, one could replace our estimate of $\dot{M}_{\ast}$ in \autoref{eqn:kappa.min} with a similar scaling proportional to the black hole accretion rate. However, the appropriate ``efficiency'' in that regime is order-of-magnitude uncertain (see \citealt{su:turb.crs.quench}). This, in turn, means that any other source of cosmic rays (for example, from accretion/virial shocks) does not change the validity of $\kappa_{\rm eff}^{\rm min}$ as a lower limit.

Still, there is one major regime where the derived scaling relation can break down. If collisional or collisionless
losses within the galaxy deplete most of the cosmic-ray energy, then the cosmic-ray pressure (\autoref{eqn:pcr.profile}) can be reduced by a factor $f_{\rm loss}\ll1$, so that $P_{\rm cr} \propto f_{\rm loss}\, E_{\rm cr,\, injected}$. If all of the cosmic-ray energy is lost in the ISM, then of course $\kappa_{\rm eff}$ can be arbitrarily small without having any effect on the CGM, and our lower limits are invalid. However, as discussed in \citet{Chan:2019} and \citet{Hopkins:2020,Hopkins:2021_transport1}, observations strongly imply that collisional losses are negligible ($f_{\rm loss} \approx 1$) in dwarf and Milky Way-mass galaxies at low redshifts \citep[see e.g.][and references therein]{strong:2010.milky.way.sub.calorimetric.by.factor.hundred,ackermann.2011:diffuse.gamma.ray.cr.profile.constraints,lacki:2011.cosmic.ray.sub.calorimetric,ackermann:2012.fermi.obs.cr.emissivity.variation,yoast.hull:2013.m82.electron.calorimeter.but.not.proton,2014ApJ...780..137Y,tibaldo.2014:diffuse.gamma.ray.cr.profile.constraints,gaggero:2015.cr.diffusion.coefficient,tibaldo.2015:diffuse.gamma.ray.cr.profile.constraints,acero:2016.gamma.ray.constraints.cr.emissivity,yang.2016:diffuse.gamma.ray.cr.profile.constraints,2016MNRAS.463.1068R,2017PhRvD..95h2007A,lopez:2018.smc.below.calorimetric.crs,2019ApJ...874..173Z,delaTorre:2021.dragon2.methods.new.model.comparison,tibaldo.2021:diffuse.gamma.ray.cr.profile.constraints,2021Ap&SS.366..117H,2022A&A...657A..49K,2022arXiv220802059P}. 
Similarly, collisionless losses are also expected to be negligible in dwarf and Milky Way-mass galaxies \citep{Hopkins:2020, Ji:2020}. Yet, cosmic-ray losses can be significant ($f_{\rm loss} \lesssim 1$) in starburst galaxies with gas surface densities $\gg 50\,M_{\odot}\,{\rm pc^{-2}}$ (star formation rate surface densities $\gtrsim 0.1\,{\rm M_{\odot}\,yr^{-1}\,kpc^{-2}}$). Nevertheless, as long as $f_{\rm loss}$ is not extremely small ($\ll 1$), i.e.\ if an order-unity fraction of the energy in $\sim 1-10$\,GeV protons can escape from galaxies, then our method should be robust.

That said, the closer a galaxy halo is to actually being dominated by cosmic-ray pressure from supernovae, the closer our lower limit will come to equaling the true $\kappa_{\rm eff}$, and so it will (all else equal) tend to be more ``interesting.'' In contrast, if the system has negligible cosmic-ray pressure, then the lower limits are more likely to give un-interesting (e.g.\ very low and therefore not particularly constraining) values. As discussed extensively in \citet{Hopkins:2020}, the fraction of cosmic-ray pressure support in the CGM is predicted to be a strong function of simulated galaxy mass, strongest in galaxies with halo masses $\sim 10^{11}-10^{12.5}\,M_{\odot}$ (stellar masses $\sim 10^{9}-10^{11}\,M_{\odot}$), at redshifts $z\lesssim 1-2$.

This suggests that our comparisons are most interesting in dwarf-through-$\lstar$, non-starbursting galaxies at $z\lesssim 1-2$.

\begin{table*}
    \centering
    \begin{tabular}{lcccc}
    \hline
    Simulation & M$^{\rm vir}_{\rm halo}$ [M$_\odot$]& M$^{\rm MHD}_{\rm *}$ [M$_\odot$] & M$^{\rm CR}_{\rm *}$ [M$_\odot$] & Description \\
    \hline
    m11i    & 6.8e10 & 6e8 & 2e8   & Dwarf with late mergers and accretion \\
    m11f    & 5.2e11 & 3e10 & 1e10 & Early forming intermediate-mass halo \\
    m12i    & 1.2e12 & 7e10 & 3e10 & Late forming MW-mass halo with a massive disk \\
    \hline
    \end{tabular}
    \caption{The virial and stellar masses at $z = 0$ of the simulations studied in this work. For additional information, we refer the reader to \citet{Hopkins:2020}. } 
    \label{tab:sims}
\end{table*}

\begin{figure*}
\begin{centering}
\includegraphics[width = \textwidth]{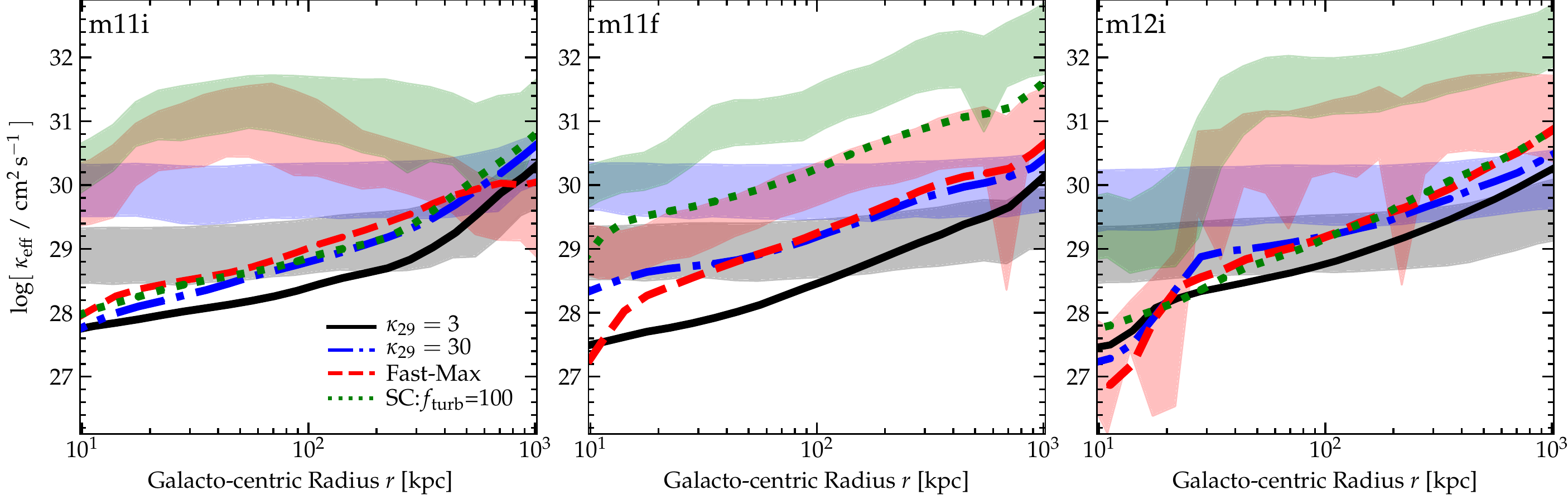}
\end{centering}
    \caption{Comparison of the true (simulation) effective cosmic-ray transport rate, $\kappa_{\rm eff}$ (shaded), and the lower limit inferred from taking \autoref{eqn:kappa.min} ($\kappa_{\rm eff}^{\rm min}$; solid, dashed, and dotted lines) for four different cosmic-ray transport models.   We compare three galaxies: a small dwarf ({\bf m11i}; {\em left}), intermediate-mass disk ({\bf m11f}; {\em middle}), and massive bulge+disk ({\bf m12i}; {\em right}) at $z = 0$. For each, we estimate the ``true'' $\kappa_{\rm eff}$ as the angle-averaged (isotropic-equivalent) effective cosmic-ray transport rate $\langle |{\bf F}_{\rm cr}\cdot \hat{\bf r}| / |\nabla e_{\rm cr} | \rangle$ weighted by the cosmic-ray scattering rate. As predicted by \autoref{eqn:kappa.min}, $\kmin \le \kappa_{\rm eff}$, is a robust lower limit within several hundred kiloparsecs of the galaxy. 
    \label{fig:demo.alt.kappa}}
\end{figure*}

\section{Validation in Numerical Simulations}
\label{sec:sims}

We now validate these scalings with a comparison to galaxy simulations from \citet{Hopkins:2020, Hopkins:2021_transport1}. These are fully cosmological, high-resolution (mass resolution $\approx 7000\,M_{\sun}$) zoom-in simulations, which include explicit treatment of radiative cooling (with a variety of processes including metal-line cooling, self-shielding, local and meta-galactic radiation fields in multi-band radiation-transport), magneto-hydrodynamics, anisotropic Spitzer-Braginskii conduction and viscosity, star formation, stellar feedback from supernovae (Ia \&\ II), stellar mass-loss (O/B and AGB) and radiative feedback (photo-ionization and photo-electric heating and radiation pressure), as well as cosmic-ray injection from supernovae (with efficiency $\epsilon_{\rm cr}= E_{\rm cr} / E_{\rm SN} = 0.1$), and explicit two-moment, fully anisotropic cosmic-ray transport (including variable streaming and/or diffusion coefficients) using a variety of different models for the effective streaming coefficients or diffusivities and cosmic-ray scatter rates. The simulations also self-consistently account for dynamical cosmic ray-gas coupling, and all the dominant loss terms (e.g. hadronic and Coulomb losses, for protons) as well as adiabatic and advective terms.

For our comparison, we consider three different simulated galaxies of different masses: Milky Way-mass ({\bf m12i}), intermediate-mass ({\bf m11f}), and a dwarf galaxy ({\bf m11i}). These galaxies, simulated with cosmic-ray feedback, are an extension the Feedback in Realistic Environments 2 (FIRE2) galaxies described in \citet{Hopkins:2018} and \citet{Hopkins:2020}. We provide a brief summary the galaxy masses and descriptions in \autoref{tab:sims}.

For each galaxy, we consider four different cosmic-ray transport models that are all constrained to reproduce the {\em same} observational constraints on cosmic rays, including all of the $\gamma$-ray observations from the Local Group, starburst galaxies, and AGN, as well as constraints from the Milky Way Solar-circle on the cosmic-ray energy density, grammage, and residence time: ``$\kappa_{29} = 3$'', ``$\kappa_{29} = 30$'', ``Fast-Max'', and ``SC:$f_{\rm turb}=100$'' described in \citet{Chan:2019} and \citet{Hopkins:2021_transport1}. In models $\kappa_{29} = 3,30$, we model cosmic-ray transport with fully-anisotropic diffusion (with a constant parallel diffusivity $\kappa_{\|} = \kappa_{29} \times10^{29}\,{\rm cm^{2}\,s^{-1}}$) and streaming (at the local \Alf\ speed). Model ``Fast-Max'' is motivated by extrinsic turbulence models of cosmic-ray transport with some terms re-scaled to increase the cosmic-ray scattering rate. Model ``SC:$f_{\rm turb}=100$'' is motivated by ``self-confinement'' (scattering by waves self-excited by cosmic-ray streaming) models of cosmic-ray transport with re-scaled damping rates to decrease the cosmic-ray scattering rate. For a more detailed description of the cosmic-ray transport models and their impact on the evolution of the simulated galaxies and their CGM, we refer the reader to \citet{Ji:2020} and \citet{Hopkins:2020, Hopkins:2021_transport2, Hopkins:2021_transport1}.

\autoref{fig:demo.alt.kappa} directly compares the ``true'' effective $\kappa_{\rm eff}$ in the simulations, measured in different radial annuli versus the lower-limit $\kappa_{\rm eff}^{\rm min}$ calculated from \autoref{eqn:kappa.min}, reconstructed as it would be observed assuming a perfect measurement of the column density $N_{\rm H}$, star formation rate averaged in the last $\sim 100$\,Myr, and $V_{\rm c}^{2}(R)$.

In the two models with constant true anisotropic diffusion coefficients ($\kappa_{29} = 3, 30$) plus streaming at the \Alf\ speed, the true $\kappa_{\rm eff}$ remains relatively constant in the inner CGM, until the streaming term begins to dominate a few hundred kiloparsecs from the galaxy center, where $\kappa_{\rm eff}$ rises. In models Fast-Max and SC:$f_{\rm turb}=100$, the true $\kappa_{\rm eff}$ is a nonlinear function of local gas properties and therefore takes on different profile shapes in the three different galaxy models. Importantly, as shown in \citet{Hopkins:2021_transport2}, the different transport models lead to wildly different cosmic-ray pressure profiles in the CGM, and not all of them feature a cosmic-ray-pressure-supported CGM. 

In models $\kappa_{29} = 3, 30$, in which the true $\kappa_{\rm eff}$ is not explicitly tracked at simulation runtime, the $\kmin$ line intersects the expected $\kappa_{\rm eff}$ region at large galactocentric radii ($r \gtrsim r_{\rm vir}$). This effect is mainly due to numerical effects of estimating $\kappa_{\rm eff}$ as well as a decline in resolution at the outskirts of the zoom-in region. However, regardless of the cosmic-ray transport model, \textit{the lower limit of the cosmic-ray transport rate derived in \autoref{eqn:kappa.min} is robust within several hundred kiloparsecs of the galaxy.} Additionally, we note that although the simulations highlighted in \autoref{fig:demo.alt.kappa} all have effective cosmic-ray transport speeds, $\kappa_{\rm eff} \gtrsim 10^{29}\, {\rm cm}^2\,{\rm s}^{-1}$, this relation holds for slower cosmic-ray transport speeds so long as cosmic-ray losses in the disk remain negligible. 

\section{Application to Observational Constraints}
\label{sec:obs}
\begin{figure}
{\centering \includegraphics[width=0.46\textwidth]{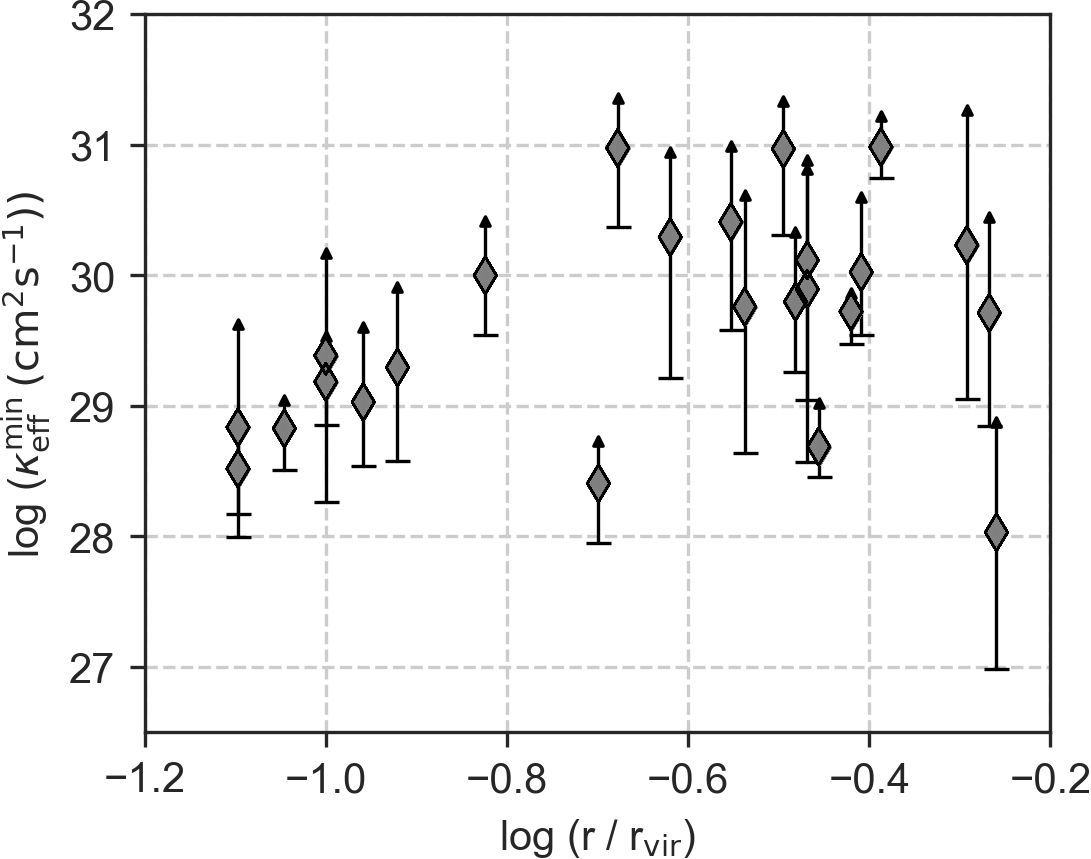}}

    \caption{The lower limit of the effective cosmic-ray transport rate (\autoref{eqn:kappa.min}) as a function of impact parameter scaled by the galactic virial radius in the star-forming galaxies from the COS-Halos survey \citep{Werk:2013, Werk:2014}. The error bars capture some of the uncertainty in the estimated total hydrogen column density ($N_{\rm H}$), average star formation rate ($\dot{M}_{\ast}$), and circular velocity ($V_{\rm c}(r)$) values that go into estimating $\kmin$. These $\kmin$ values derived from observations hint at a spatially-varying cosmic-ray transport rate in the CGM and rule out a constant $\kappa_{\rm eff} \approx 10^{28-29}\,{\rm cm}^2\,{\rm s}^{-1}$ or any $\kappa_{\rm eff} \ll 10^{30}\,{\rm cm}^{2}\,{\rm s^{-1}}$ model of cosmic-ray transport in the outer CGM ($\gtrsim 50-100$ kpc). Recall that this is isotropically-averaged: the true anisotropic $\kappa_{\|}$ will typically be a factor $\sim 3$ larger.}
    \label{fig:obs}
\end{figure}

Next, we apply our proposed scaling relation to the star-forming galaxies in the COS-Halos sample \citep{Werk:2013, Werk:2014} that have estimates for the total hydrogen column density, $N_{\rm H}$. The sample contains 24 galaxies at redshifts $0.14 < z < 0.36$ with stellar masses between $9.6 < {\rm log}(M_{\ast}/M_{\odot}) < 11.1$. These galaxies all appear to lie near or below the star formation ``main sequence'' and none meets the criterion for classification as a starburst galaxy or ULIRG (where our method is less reliable).

We take the estimates for $N_{\rm H}$ and its uncertainty directly from \citet{Werk:2014} and estimate the average star formation rate using two approaches. The first is simply using the stated current star formation rates in \citet{Werk:2013}, derived from a combination of emission-line spectroscopy and broadband photometry. This approach gives a relatively recent ($< 1$ Gyr) probe of the galaxy star formation rate, so we compensate by calculating a second, cosmic-time-averaged star formation rate by simply dividing each galaxy's total stellar mass by its age at the observed redshift). The final star formation rate value we use in \autoref{fig:obs} is the average of these two methods and the error bars bracket the range we obtain. 

Finally, we obtain estimates for the circular velocity at a given impact parameter using two different methods. In the first method, we estimate the virial mass of each galaxy from its stated stellar mass using the \citet{Moster:2010} analytic relation, and then use those virial masses to estimate analytic Navarro-Frenk-White \citep[NFW;][]{Navarro:1997} velocity profiles. In the second method, we use each galaxy’s luminosity to make an analytic estimate of its maximum circular velocity following the analytic relation for star-forming galaxies from \citet{Reyes:2011}. Again, the final $V_{\rm c}$ value used in \autoref{eqn:kappa.min} is the average of the $V_{\rm c}$ derived from these two methods and the error bars bracket the range. Overall, we find that the uncertainty in estimating the star formation rate and the circular velocity is small relative to the uncertainty in estimating $N_{\rm H}$. To be as conservative as possible, the final error bars include the estimated (measurement and systematic) uncertainty in $N_{\rm H}$ as well as the entire possible range we could obtain adopting any combination of the different $\dot{M}_{\ast}$ and/or $V_{\rm c}$ estimators defined above.

\autoref{fig:obs} shows the inferred minimum cosmic-ray transport rates from the star-forming COS-Halos galaxies as a function of their impact parameter, normalized by each galaxy’s virial radius. The inner CGM shows a notable increasing trend in $\kmin$, while the outer CGM tends to have higher $\kmin$ values with a large scatter. We note that although we attempt to capture the inherent systematic and measurement uncertainty in the $\kmin$ estimate with the error bars in the plot, they could still represent an underestimate of the true uncertainty.

Recall that the points in \autoref{fig:obs} all represent the \textit{lower limit} of the effective cosmic-ray transport rate. Therefore, the data is technically consistent with the hypothetical (though perhaps, uninteresting) case where the effective transport rate is some constant value $\kappa_{\rm eff} > 10^{31}\, {\rm cm}^2\,{\rm s}^{-1}$ throughout the halo. However, given the apparent increasing $\kmin$ trend in the inner CGM, the data hint at a spatially-varying cosmic-ray transport rate within the CGM. This implied qualitative trend is unsurprising since most of the known processes that drive cosmic-ray scattering\footnote{In most theoretical models, the cosmic-ray scattering rate is inversely related to the effective cosmic-ray transport rate.} (e.g., the strength of turbulence, magnetic field amplitude, electron densities, cosmic-ray energy densities driving streaming instabilities) become weaker in the diffuse environments typically found in the CGM and intergalactic medium \citep[see the discussion in][]{Hopkins:2021_transport2}. 

The quantitative constraint presented in \autoref{fig:obs} is stronger, suggesting that the effective cosmic-ray transport rates rise to $\kappa_{\rm eff} > 10^{31}\, {\rm cm}^2\,{\rm s}^{-1}$ for $r \gtrsim 0.25\, r_{\rm vir}$. For the $\lstar$ galaxies in the COS-Halos sample, this corresponds to physical distances of $r \gtrsim 50-100$ kpc from the galactic center. 
If other galaxies have similar cosmic-ray transport physics to those in the Milky Way ($\kappa_{\rm eff}  \sim 10^{28-29}\, {\rm cm}^2\,{\rm s}^{-1}$ within the galactic disk; e.g.,  \citealt{Trotta:2011,maurin:2018.cr.sam.favored.parameters.close.to.fire,delaTorre:2021.dragon2.methods.new.model.comparison,hopkins:cr.multibin.mw.comparison,korsmeier:2021.light.element.requires.halo.but.upper.limit.unconfined}),\textit{ then these data necessarily rule out a constant cosmic-ray transport rate between the galaxy disk and halo. }

As noted above, the effective cosmic-ray transport rate can arise from diffusive or streaming type behavior. The minimum effective field-parallel cosmic-ray streaming speed, correcting for geometry (from $\kmin$ being isotropically averaged), implied by the lower-limit $\kappa_{\rm eff} > 10^{31}\, {\rm cm}^2\,{\rm s}^{-1}$ at $\sim 100$\,kpc becomes $v_{\rm eff}^{\rm stream} \gtrsim 1000\,{\rm km\,s^{-1}}$ in the outer CGM. Therefore, regardless of whether the constraint is phrased in terms of streaming speed, diffusivity, or advection with galactic outflows, we obtain robust indications that cosmic-ray transport speeds increase in the halo and outer CGM compared to standard ISM values.

\section{Conclusions}
\label{sec:conclusions}

We present a simple analytic model that provides a robust lower limit to the effective cosmic-ray transport rate for the $\sim 1-10$\,GeV protons that dominate the cosmic-ray energy density and pressure in the circumgalactic and intergalactic medium (CGM/IGM), where thus far almost no observational constraints exist. Such constraints are of profound importance, not only for informing models of cosmic-ray transport themselves, but also for models of how cosmic rays influence galaxy formation through, e.g., the generation of galactic winds, cooling flows, and the phase structure of the CGM/IGM. We demonstrate that as long as a non-negligible fraction of cosmic rays escape the galactic disk, then we can estimate a minimum effective cosmic-ray transport rate, $\kmin$, that is a function of three observed quantities: the galaxy star-formation rate ($\dot{M}_{\ast}$; ideally averaged over relatively long timescales), the circular velocity ($V_{\rm c}(r)$), and the total hydrogen column density ($N_{\rm H}$). We note that $\kmin$ describes the \textit{total effective} cosmic-ray transport rate, not only the transport rate relative to the gas, and is agnostic to the cosmic-ray transport model (e.g., diffusion, streaming, or advection).

We validate this relation and the underlying assumptions using a suite of simulated galaxies with a range of masses and cosmic-ray transport physics, all calibrated to reproduce existing cosmic-ray observational constraints. Although some of these simulations have CGM that are dominated by cosmic-ray pressure and others have halos with negligibly small cosmic-ray pressures, all validate our method as a robust lower limit. We also show the lower limit remains remains a valid lower limit even if other sources of cosmic rays (e.g.\ from AGN or structure formation shocks) dominate. We caution against application to extreme starburst galaxies like Arp 220, where cosmic-ray protons suffer significant losses in the galactic disk. However, excluding these systems, we show that for dwarf-through-$\sim \lstar$, star-forming galaxies at $z\lesssim 1-2$, this simple relation for $\kmin$ (\autoref{eqn:kappa.min}) is a robust lower limit to the true cosmic-ray transport rate (\autoref{fig:demo.alt.kappa}). 

We then apply our model to estimate $\kmin$ for the star-forming COS-Halos galaxies in \autoref{fig:obs}. Even with large observational systematic uncertainties in the input values $N_{\rm H}$, $V_{\rm c}(r)$, and $\dot{M}_{\ast}$, the fact that there are, at present, essentially {\em no} constraints on the cosmic-ray transport coefficients for $\sim$\,GeV protons in the CGM/IGM means that essentially every one of these limits, even with its attendant uncertainty, is extremely interesting. We find that the resulting $\kmin$ values strongly suggest a spatially-varying $\kappa_{\rm eff}$ in the CGM, rising from the canonical observationally-constrained ISM values at $\lesssim 10\,$kpc to values $\kappa_{\rm eff} \gtrsim 10^{30-31}\,{\rm cm^{2}\,s^{-1}}$ ($v_{\rm eff}^{\rm stream} \gtrsim 1000\,{\rm km\,s^{-1}}$) in the outer halo at galactocentric distances $\gtrsim 50-100\,$kpc. The data appear to quite clearly rule out any model where the cosmic-ray transport rate is constant, with values $\kappa_{\rm eff} \ll 10^{30}\, {\rm cm}^2\,{\rm s}^{-1}$ in the CGM.

There are, of course, more detailed physical uncertainties underpinning the assumptions made here, which could be refined with additional observational data to sharpen the constraints presented here. Physically, we do assume {\em some} cosmic-ray gas coupling so that cosmic-ray pressure could, in principle act on gas in the CGM. One might then wonder, if cosmic-ray scattering were to somehow cease in the CGM, if it would render our conclusions invalid. Since the zero-scattering limit is formally equivalent to the limit where $\kappa_{\rm eff} \rightarrow \infty$ (cosmic rays free-stream at $c$), our lower limits would remain valid lower limits. However, this limit is not expected in any reasonable physical model. 

Existing observational constraints on cosmic rays, including $\gamma$-ray emission, grammage, residence times, energy densities and ionization rates, all only constrain cosmic rays around the star-forming ISM (at $R \lesssim 10\,$kpc). As a result, different theoretical cosmic-ray transport models can reproduce these same observations with similar $\kappa_{\rm eff}$ in the ISM, but with $\kappa_{\rm eff}$ diverging by up to $\sim 4$\,dex in the CGM, where there are essentially no constraints at the present (at least for the typical dwarf and Milky Way-mass galaxies in the sample considered here). Thus, the proposed constraint on $\kmin$ in the CGM presents a {\em qualitative} change to our ability to understand cosmic-ray microphysics and, therefore, the role cosmic rays may play in galaxy formation and plasma physics in galaxy halos.

\acknowledgments 
The authors would like to thank Gina Panopoulou for her insightful comments on this work. I.S.B. was supported by the DuBridge Postdoctoral Fellowship at Caltech. Support for P.F.H. and co-authors was provided by NSF Research Grants 1911233 \&\ 20009234, NSF CAREER grant 1455342, NASA grants 80NSSC18K0562, HST-AR-15800.001-A. Numerical calculations were run on the Caltech compute cluster ``Wheeler,'' allocations FTA-Hopkins supported by the NSF and TACC, and NASA HEC SMD-16-7592. The data used in this work were, in part, hosted on facilities supported by the Scientific Computing Core at the Flatiron Institute, a division of the Simons Foundation.

\datastatement{The data supporting this article are available on reasonable request to the corresponding author. A public version of the
GIZMO code is available at http://www.tapir.caltech.
edu/~phopkins/Site/GIZMO.html.}
\\
\\
\\

\bibliography{main}

\begin{thebibliography}{}
\expandafter\ifx\csname natexlab\endcsname\relax\def\natexlab#1{#1}\fi
\providecommand{\url}[1]{\href{#1}{#1}}
\providecommand{\dodoi}[1]{doi:~\href{http://doi.org/#1}{\nolinkurl{#1}}}
\providecommand{\doeprint}[1]{\href{http://ascl.net/#1}{\nolinkurl{http://ascl.net/#1}}}
\providecommand{\doarXiv}[1]{\href{https://arxiv.org/abs/#1}{\nolinkurl{https://arxiv.org/abs/#1}}}

\bibitem[{{Abdollahi} {et~al.}(2017){Abdollahi}, {Ackermann}, {Ajello},
  {Atwood}, {Baldini}, {Barbiellini}, {Bastieri}, {Bellazzini}, {Bloom},
  {Bonino}, {Brandt}, {Bregeon}, {Bruel}, {Buehler}, {Cameron}, {Caputo},
  {Caragiulo}, {Castro}, {Cavazzuti}, {Cecchi}, {Chekhtman}, {Ciprini},
  {Cohen-Tanugi}, {Costanza}, {Cuoco}, {Cutini}, {D'Ammando}, {de Palma},
  {Desiante}, {Digel}, {Di Lalla}, {Di Mauro}, {Di Venere}, {Drell},
  {Drlica-Wagner}, {Favuzzi}, {Focke}, {Funk}, {Fusco}, {Gargano},
  {Gasparrini}, {Giglietto}, {Giordano}, {Giroletti}, {Green}, {Guillemot},
  {Guiriec}, {Harding}, {Jogler}, {J{\'o}hannesson}, {Kamae}, {Kuss}, {La
  Mura}, {Latronico}, {Longo}, {Loparco}, {Lubrano}, {Maldera}, {Malyshev},
  {Manfreda}, {Mazziotta}, {Michelson}, {Mirabal}, {Mitthumsiri}, {Mizuno},
  {Moiseev}, {Monzani}, {Morselli}, {Moskalenko}, {Negro}, {Nuss}, {Orlando},
  {Paneque}, {Perkins}, {Pesce-Rollins}, {Piron}, {Pivato}, {Porter},
  {Principe}, {Rain{\`o}}, {Rando}, {Razzano}, {Reimer}, {Reimer}, {Sgr{\`o}},
  {Simone}, {Siskind}, {Spada}, {Spandre}, {Spinelli}, {Tajima}, {Thayer},
  {Tibaldo}, {Torres}, {Troja}, {Wood}, {Worley}, {Zaharijas}, {Zimmer}, \&
  {Fermi-LAT Collaboration}}]{2017PhRvD..95h2007A}
{Abdollahi}, S., {Ackermann}, M., {Ajello}, M., {et~al.} 2017, \prd, 95,
  082007, \dodoi{10.1103/PhysRevD.95.082007}

\bibitem[{{Acero} {et~al.}(2016){Acero}, {Ackermann}, {Ajello}, {Albert},
  {Baldini}, {Ballet}, {Barbiellini}, {Bastieri}, {Bellazzini}, {Bissaldi},
  {Bloom}, {Bonino}, {Bottacini}, {Brandt}, {Bregeon}, {Bruel}, {Buehler},
  {Buson}, {Caliandro}, {Cameron}, {Caragiulo}, {Caraveo}, {Casandjian},
  {Cavazzuti}, {Cecchi}, {Charles}, {Chekhtman}, {Chiang}, {Chiaro}, {Ciprini},
  {Claus}, {Cohen-Tanugi}, {Conrad}, {Cuoco}, {Cutini}, {D'Ammando}, {de
  Angelis}, {de Palma}, {Desiante}, {Digel}, {Di Venere}, {Drell}, {Favuzzi},
  {Fegan}, {Ferrara}, {Focke}, {Franckowiak}, {Funk}, {Fusco}, {Gargano},
  {Gasparrini}, {Giglietto}, {Giordano}, {Giroletti}, {Glanzman}, {Godfrey},
  {Grenier}, {Guiriec}, {Hadasch}, {Harding}, {Hayashi}, {Hays}, {Hewitt},
  {Hill}, {Horan}, {Hou}, {Jogler}, {J{\'o}hannesson}, {Kamae}, {Kuss},
  {Landriu}, {Larsson}, {Latronico}, {Li}, {Li}, {Longo}, {Loparco},
  {Lovellette}, {Lubrano}, {Maldera}, {Malyshev}, {Manfreda}, {Martin},
  {Mayer}, {Mazziotta}, {McEnery}, {Michelson}, {Mirabal}, {Mizuno}, {Monzani},
  {Morselli}, {Nuss}, {Ohsugi}, {Omodei}, {Orienti}, {Orlando}, {Ormes},
  {Paneque}, {Pesce-Rollins}, {Piron}, {Pivato}, {Rain{\`o}}, {Rando},
  {Razzano}, {Razzaque}, {Reimer}, {Reimer}, {Remy}, {Renault},
  {S{\'a}nchez-Conde}, {Schaal}, {Schulz}, {Sgr{\`o}}, {Siskind}, {Spada},
  {Spandre}, {Spinelli}, {Strong}, {Suson}, {Tajima}, {Takahashi}, {Thayer},
  {Thompson}, {Tibaldo}, {Tinivella}, {Torres}, {Tosti}, {Troja}, {Vianello},
  {Werner}, {Wood}, {Wood}, {Zaharijas}, \&
  {Zimmer}}]{acero:2016.gamma.ray.constraints.cr.emissivity}
{Acero}, F., {Ackermann}, M., {Ajello}, M., {et~al.} 2016, \apjs, 223, 26,
  \dodoi{10.3847/0067-0049/223/2/26}

\bibitem[{{Ackermann} {et~al.}(2011){Ackermann}, {Ajello}, {Baldini}, {Ballet},
  {Barbiellini}, {Bastieri}, {Bechtol}, {Bellazzini}, {Berenji}, {Bloom},
  {Bonamente}, {Borgland}, {Brandt}, {Bregeon}, {Brez}, {Brigida}, {Bruel},
  {Buehler}, {Buson}, {Caliandro}, {Cameron}, {Caraveo}, {Casandjian},
  {Cecchi}, {Charles}, {Chekhtman}, {Chiang}, {Ciprini}, {Claus},
  {Cohen-Tanugi}, {Conrad}, {Dermer}, {de Palma}, {Digel}, {Drell}, {Dubois},
  {Favuzzi}, {Ferrara}, {Focke}, {Fukazawa}, {Funk}, {Fusco}, {Gargano},
  {Germani}, {Giglietto}, {Giordano}, {Giroletti}, {Glanzman}, {Godfrey},
  {Grenier}, {Guiriec}, {Hadasch}, {Hanabata}, {Harding}, {Hayashi},
  {Hayashida}, {Hughes}, {Itoh}, {J{\'o}hannesson}, {Johnson}, {Johnson},
  {Kamae}, {Katagiri}, {Kataoka}, {Kn{\"o}dlseder}, {Kuss}, {Lande},
  {Latronico}, {Lee}, {Llena Garde}, {Longo}, {Loparco}, {Lovellette},
  {Lubrano}, {Makeev}, {Martin}, {Mazziotta}, {McEnery}, {Mehault},
  {Michelson}, {Mizuno}, {Monte}, {Monzani}, {Morselli}, {Moskalenko},
  {Murgia}, {Naumann-Godo}, {Nishino}, {Nolan}, {Norris}, {Nuss}, {Ohsugi},
  {Okumura}, {Omodei}, {Orlando}, {Ormes}, {Ozaki}, {Parent}, {Pelassa},
  {Pepe}, {Pesce-Rollins}, {Piron}, {Porter}, {Rain{\`o}}, {Rando}, {Razzano},
  {Reimer}, {Reimer}, {Ripken}, {Sada}, {Sadrozinski}, {Sgr{\`o}}, {Siskind},
  {Spandre}, {Spinelli}, {Strickman}, {Strong}, {Suson}, {Takahashi},
  {Takahashi}, {Tanaka}, {Thayer}, {Thompson}, {Tibaldo}, {Torres},
  {Tramacere}, {Uchiyama}, {Uehara}, {Usher}, {Vandenbroucke}, {Vasileiou},
  {Vilchez}, {Vitale}, {Vladimirov}, {Waite}, {Wang}, {Wood}, {Yang}, \&
  {Ziegler}}]{ackermann.2011:diffuse.gamma.ray.cr.profile.constraints}
{Ackermann}, M., {Ajello}, M., {Baldini}, L., {et~al.} 2011, \apj, 726, 81,
  \dodoi{10.1088/0004-637X/726/2/81}

\bibitem[{{Ackermann} {et~al.}(2012){Ackermann}, {Ajello}, {Allafort},
  {Baldini}, {Ballet}, {Barbiellini}, {Bastieri}, {Bechtol}, {Bellazzini},
  {Berenji}, {Blandford}, {Bloom}, {Bonamente}, {Borgland}, {Bottacini},
  {Brandt}, {Bregeon}, {Brigida}, {Bruel}, {Buehler}, {Busetto}, {Buson},
  {Caliandro}, {Cameron}, {Caraveo}, {Casandjian}, {Cecchi}, {Charles},
  {Chekhtman}, {Chiang}, {Ciprini}, {Claus}, {Cohen-Tanugi}, {Conrad},
  {D'Ammando}, {de Angelis}, {de Palma}, {Dermer}, {Digel}, {Silva}, {Drell},
  {Drlica-Wagner}, {Falletti}, {Favuzzi}, {Fegan}, {Ferrara}, {Focke},
  {Fukazawa}, {Fukui}, {Funk}, {Fusco}, {Gargano}, {Gasparrini}, {Germani},
  {Giglietto}, {Giordano}, {Giroletti}, {Glanzman}, {Godfrey}, {Grenier},
  {Grondin}, {Grove}, {Guiriec}, {Hadasch}, {Hanabata}, {Harding}, {Hayashi},
  {Horan}, {Hou}, {Hughes}, {Itoh}, {Jackson}, {J{\'o}hannesson}, {Johnson},
  {Kamae}, {Katagiri}, {Kataoka}, {Kn{\"o}dlseder}, {Kuss}, {Lande}, {Larsson},
  {Lee}, {Lemoine-Goumard}, {Longo}, {Loparco}, {Lovellette}, {Lubrano},
  {Martin}, {Mazziotta}, {McEnery}, {Mehault}, {Michelson}, {Mitthumsiri},
  {Mizuno}, {Moiseev}, {Monte}, {Monzani}, {Morselli}, {Moskalenko}, {Murgia},
  {Naumann-Godo}, {Nemmen}, {Nishino}, {Norris}, {Nuss}, {Ohno}, {Ohsugi},
  {Okumura}, {Omodei}, {Orlando}, {Ormes}, {Ozaki}, {Paneque}, {Panetta},
  {Parent}, {Pesce-Rollins}, {Pierbattista}, {Piron}, {Pivato}, {Porter},
  {Rain{\`o}}, {Rando}, {Razzano}, {Reimer}, {Reimer}, {Romoli}, {Roth},
  {Sada}, {Sadrozinski}, {Sanchez}, {Sbarra}, {Sgr{\`o}}, {Siskind}, {Spandre},
  {Spinelli}, {Strong}, {Suson}, {Takahashi}, {Takahashi}, {Tanaka}, {Thayer},
  {Thayer}, {Thompson}, {Tibaldo}, {Tibolla}, {Tinivella}, {Torres}, {Tosti},
  {Tramacere}, {Troja}, {Uchiyama}, {Uehara}, {Usher}, {Vandenbroucke},
  {Vasileiou}, {Vianello}, {Vitale}, {Waite}, {Wang}, {Winer}, {Wood},
  {Yamamoto}, {Yang}, \&
  {Zimmer}}]{ackermann:2012.fermi.obs.cr.emissivity.variation}
{Ackermann}, M., {Ajello}, M., {Allafort}, A., {et~al.} 2012, \apj, 755, 22,
  \dodoi{10.1088/0004-637X/755/1/22}

\bibitem[{{Booth} {et~al.}(2013){Booth}, {Agertz}, {Kravtsov}, \&
  {Gnedin}}]{Booth:2013}
{Booth}, C.~M., {Agertz}, O., {Kravtsov}, A.~V., \& {Gnedin}, N.~Y. 2013,
  \apjl, 777, L16, \dodoi{10.1088/2041-8205/777/1/L16}

\bibitem[{{Buck} {et~al.}(2020){Buck}, {Pfrommer}, {Pakmor}, {Grand}, \&
  {Springel}}]{Buck:2020}
{Buck}, T., {Pfrommer}, C., {Pakmor}, R., {Grand}, R. J.~J., \& {Springel}, V.
  2020, \mnras, 497, 1712, \dodoi{10.1093/mnras/staa1960}

\bibitem[{{Bustard} {et~al.}(2020){Bustard}, {Zweibel}, {D'Onghia},
  {Gallagher}, \& {Farber}}]{Bustard:2020}
{Bustard}, C., {Zweibel}, E.~G., {D'Onghia}, E., {Gallagher}, J.~S., I., \&
  {Farber}, R. 2020, \apj, 893, 29, \dodoi{10.3847/1538-4357/ab7fa3}

\bibitem[{Butsky {et~al.}(2020)Butsky, Fielding, Hayward, Hummels, Quinn, \&
  Werk}]{Butsky:2020}
Butsky, I.~S., Fielding, D.~B., Hayward, C.~C., {et~al.} 2020, \apj, 903, 77,
  \dodoi{10.3847/1538-4357/abbad2}

\bibitem[{{Butsky} \& {Quinn}(2018)}]{Butsky:2018}
{Butsky}, I.~S., \& {Quinn}, T.~R. 2018, \apj, 868, 108,
  \dodoi{10.3847/1538-4357/aaeac2}

\bibitem[{{Butsky} {et~al.}(2022){Butsky}, {Werk}, {Tchernyshyov}, {Fielding},
  {Breneman}, {Piacitelli}, {Quinn}, {Sanchez}, {Cruz}, {Hummels}, {Burchett},
  \& {Tremmel}}]{Butsky:2022}
{Butsky}, I.~S., {Werk}, J.~K., {Tchernyshyov}, K., {et~al.} 2022, \apj, 935,
  69, \dodoi{10.3847/1538-4357/ac7ebd}

\bibitem[{{Chan} {et~al.}(2022){Chan}, {Kere{\v{s}}}, {Gurvich}, {Hopkins},
  {Trapp}, {Ji}, \& {Faucher-Gigu{\`e}re}}]{Chan:2022}
{Chan}, T.~K., {Kere{\v{s}}}, D., {Gurvich}, A.~B., {et~al.} 2022, \mnras, 517,
  597, \dodoi{10.1093/mnras/stac2236}

\bibitem[{{Chan} {et~al.}(2019){Chan}, {Kere{\v{s}}}, {Hopkins}, {Quataert},
  {Su}, {Hayward}, \& {Faucher-Gigu{\`e}re}}]{Chan:2019}
{Chan}, T.~K., {Kere{\v{s}}}, D., {Hopkins}, P.~F., {et~al.} 2019, \mnras, 488,
  3716, \dodoi{10.1093/mnras/stz1895}

\bibitem[{{De La Torre Luque} {et~al.}(2021){De La Torre Luque}, {Mazziotta},
  {Loparco}, {Gargano}, \&
  {Serini}}]{delaTorre:2021.dragon2.methods.new.model.comparison}
{De La Torre Luque}, P., {Mazziotta}, M.~N., {Loparco}, F., {Gargano}, F., \&
  {Serini}, D. 2021, \jcap, 2021, 099, \dodoi{10.1088/1475-7516/2021/03/099}

\bibitem[{{Do} {et~al.}(2021){Do}, {Duong}, {McDaniel}, {O'Connor}, {Profumo},
  {Rafael}, {Sweeney}, \& {Vera}}]{Do:2021}
{Do}, A., {Duong}, M., {McDaniel}, A., {et~al.} 2021, \prd, 104, 123016,
  \dodoi{10.1103/PhysRevD.104.123016}

\bibitem[{{Farber} {et~al.}(2018){Farber}, {Ruszkowski}, {Yang}, \&
  {Zweibel}}]{Farber:2018}
{Farber}, R., {Ruszkowski}, M., {Yang}, H. Y.~K., \& {Zweibel}, E.~G. 2018,
  \apj, 856, 112, \dodoi{10.3847/1538-4357/aab26d}

\bibitem[{{Feldmann} {et~al.}(2013){Feldmann}, {Hooper}, \&
  {Gnedin}}]{Feldmann:2013}
{Feldmann}, R., {Hooper}, D., \& {Gnedin}, N.~Y. 2013, \apj, 763, 21,
  \dodoi{10.1088/0004-637X/763/1/21}

\bibitem[{{Gabici} {et~al.}(2019){Gabici}, {Evoli}, {Gaggero}, {Lipari},
  {Mertsch}, {Orlando}, {Strong}, \& {Vittino}}]{Gabici:2019}
{Gabici}, S., {Evoli}, C., {Gaggero}, D., {et~al.} 2019, International Journal
  of Modern Physics D, 28, 1930022, \dodoi{10.1142/S0218271819300222}

\bibitem[{{Gaggero} {et~al.}(2015){Gaggero}, {Urbano}, {Valli}, \&
  {Ullio}}]{gaggero:2015.cr.diffusion.coefficient}
{Gaggero}, D., {Urbano}, A., {Valli}, M., \& {Ullio}, P. 2015, \prd, 91,
  083012, \dodoi{10.1103/PhysRevD.91.083012}

\bibitem[{{Ginzburg} \& {Syrovatskii}(1964)}]{Ginzburg:1964}
{Ginzburg}, V.~L., \& {Syrovatskii}, S.~I. 1964, {The Origin of Cosmic Rays}

\bibitem[{{Girichidis} {et~al.}(2018){Girichidis}, {Naab}, {Hanasz}, \&
  {Walch}}]{Girichidis:2018}
{Girichidis}, P., {Naab}, T., {Hanasz}, M., \& {Walch}, S. 2018, \mnras, 479,
  3042, \dodoi{10.1093/mnras/sty1653}

\bibitem[{{Gupta} {et~al.}(2021){Gupta}, {Sharma}, \& {Mignone}}]{Gupta:2021}
{Gupta}, S., {Sharma}, P., \& {Mignone}, A. 2021, \mnras, 502, 2733,
  \dodoi{10.1093/mnras/stab142}

\bibitem[{{Heesen}(2021)}]{2021Ap&SS.366..117H}
{Heesen}, V. 2021, \apss, 366, 117, \dodoi{10.1007/s10509-021-04026-1}

\bibitem[{{Hopkins} {et~al.}(2022{\natexlab{a}}){Hopkins}, {Butsky}, \&
  {Ji}}]{Hopkins:2022_subgrid}
{Hopkins}, P.~F., {Butsky}, I.~S., \& {Ji}, S. 2022{\natexlab{a}}, arXiv
  e-prints, arXiv:2211.05811, \dodoi{10.48550/arXiv.2211.05811}

\bibitem[{{Hopkins} {et~al.}(2022{\natexlab{b}}){Hopkins}, {Butsky},
  {Panopoulou}, {Ji}, {Quataert}, {Faucher-Gigu{\`e}re}, \&
  {Kere{\v{s}}}}]{hopkins:cr.multibin.mw.comparison}
{Hopkins}, P.~F., {Butsky}, I.~S., {Panopoulou}, G.~V., {et~al.}
  2022{\natexlab{b}}, \mnras, 516, 3470, \dodoi{10.1093/mnras/stac1791}

\bibitem[{{Hopkins} {et~al.}(2021{\natexlab{a}}){Hopkins}, {Chan}, {Ji},
  {Hummels}, {Kere{\v{s}}}, {Quataert}, \&
  {Faucher-Gigu{\`e}re}}]{Hopkins:2021_outflows}
{Hopkins}, P.~F., {Chan}, T.~K., {Ji}, S., {et~al.} 2021{\natexlab{a}}, \mnras,
  501, 3640, \dodoi{10.1093/mnras/staa3690}

\bibitem[{{Hopkins} {et~al.}(2021{\natexlab{b}}){Hopkins}, {Chan}, {Squire},
  {Quataert}, {Ji}, {Kere{\v{s}}}, \&
  {Faucher-Gigu{\`e}re}}]{Hopkins:2021_transport2}
{Hopkins}, P.~F., {Chan}, T.~K., {Squire}, J., {et~al.} 2021{\natexlab{b}},
  \mnras, 501, 3663, \dodoi{10.1093/mnras/staa3692}

\bibitem[{{Hopkins} {et~al.}(2022{\natexlab{c}}){Hopkins}, {Squire}, {Butsky},
  \& {Ji}}]{hopkins:2021.sc.et.models.incompatible.obs}
{Hopkins}, P.~F., {Squire}, J., {Butsky}, I.~S., \& {Ji}, S.
  2022{\natexlab{c}}, \mnras, \dodoi{10.1093/mnras/stac2909}

\bibitem[{{Hopkins} {et~al.}(2021{\natexlab{c}}){Hopkins}, {Squire}, {Chan},
  {Quataert}, {Ji}, {Kere{\v{s}}}, \&
  {Faucher-Gigu{\`e}re}}]{Hopkins:2021_transport1}
{Hopkins}, P.~F., {Squire}, J., {Chan}, T.~K., {et~al.} 2021{\natexlab{c}},
  \mnras, 501, 4184, \dodoi{10.1093/mnras/staa3691}

\bibitem[{{Hopkins} {et~al.}(2018){Hopkins}, {Wetzel}, {Kere{\v s}},
  {Faucher-Gigu{\`e}re}, {Quataert}, {Boylan-Kolchin}, {Murray}, {Hayward}, \&
  {El-Badry}}]{Hopkins:2018}
{Hopkins}, P.~F., {Wetzel}, A., {Kere{\v s}}, D., {et~al.} 2018, \mnras, 477,
  1578, \dodoi{10.1093/mnras/sty674}

\bibitem[{{Hopkins} {et~al.}(2020){Hopkins}, {Chan}, {Garrison-Kimmel}, {Ji},
  {Su}, {Hummels}, {Kere{\v{s}}}, {Quataert}, \&
  {Faucher-Gigu{\`e}re}}]{Hopkins:2020}
{Hopkins}, P.~F., {Chan}, T.~K., {Garrison-Kimmel}, S., {et~al.} 2020, \mnras,
  492, 3465, \dodoi{10.1093/mnras/stz3321}

\bibitem[{{Jana} {et~al.}(2020){Jana}, {Roy}, \& {Nath}}]{Jana:2020}
{Jana}, R., {Roy}, M., \& {Nath}, B.~B. 2020, \apjl, 903, L9,
  \dodoi{10.3847/2041-8213/abbee4}

\bibitem[{{Ji} {et~al.}(2021){Ji}, {Kere{\v{s}}}, {Chan}, {Stern}, {Hummels},
  {Hopkins}, {Quataert}, \& {Faucher-Gigu{\`e}re}}]{Ji:2021}
{Ji}, S., {Kere{\v{s}}}, D., {Chan}, T.~K., {et~al.} 2021, \mnras, 505, 259,
  \dodoi{10.1093/mnras/stab1264}

\bibitem[{{Ji} {et~al.}(2020){Ji}, {Chan}, {Hummels}, {Hopkins}, {Stern},
  {Kere{\v{s}}}, {Quataert}, {Faucher-Gigu{\`e}re}, \& {Murray}}]{Ji:2020}
{Ji}, S., {Chan}, T.~K., {Hummels}, C.~B., {et~al.} 2020, \mnras,
  \dodoi{10.1093/mnras/staa1849}

\bibitem[{{Kornecki} {et~al.}(2022){Kornecki}, {Peretti}, {del Palacio},
  {Benaglia}, \& {Pellizza}}]{2022A&A...657A..49K}
{Kornecki}, P., {Peretti}, E., {del Palacio}, S., {Benaglia}, P., \&
  {Pellizza}, L.~J. 2022, \aap, 657, A49, \dodoi{10.1051/0004-6361/202141295}

\bibitem[{{Korsmeier} \&
  {Cuoco}(2021)}]{korsmeier:2021.light.element.requires.halo.but.upper.limit.unconfined}
{Korsmeier}, M., \& {Cuoco}, A. 2021, \prd, 103, 103016,
  \dodoi{10.1103/PhysRevD.103.103016}

\bibitem[{{Lacki} {et~al.}(2011){Lacki}, {Thompson}, {Quataert}, {Loeb}, \&
  {Waxman}}]{lacki:2011.cosmic.ray.sub.calorimetric}
{Lacki}, B.~C., {Thompson}, T.~A., {Quataert}, E., {Loeb}, A., \& {Waxman}, E.
  2011, \apj, 734, 107, \dodoi{10.1088/0004-637X/734/2/107}

\bibitem[{{Lopez} {et~al.}(2018){Lopez}, {Auchettl}, {Linden}, {Bolatto},
  {Thompson}, \& {Ramirez-Ruiz}}]{lopez:2018.smc.below.calorimetric.crs}
{Lopez}, L.~A., {Auchettl}, K., {Linden}, T., {et~al.} 2018, \apj, 867, 44,
  \dodoi{10.3847/1538-4357/aae0f8}

\bibitem[{{Maurin}(2020)}]{maurin:2018.cr.sam.favored.parameters.close.to.fire}
{Maurin}, D. 2020, Computer Physics Communications, 247, 106942,
  \dodoi{10.1016/j.cpc.2019.106942}

\bibitem[{{Moster} {et~al.}(2010){Moster}, {Somerville}, {Maulbetsch}, {van den
  Bosch}, {Macci{\`o}}, {Naab}, \& {Oser}}]{Moster:2010}
{Moster}, B.~P., {Somerville}, R.~S., {Maulbetsch}, C., {et~al.} 2010, \apj,
  710, 903, \dodoi{10.1088/0004-637X/710/2/903}

\bibitem[{{Navarro} {et~al.}(1997){Navarro}, {Frenk}, \&
  {White}}]{Navarro:1997}
{Navarro}, J.~F., {Frenk}, C.~S., \& {White}, S.~D.~M. 1997, \apj, 490, 493

\bibitem[{{Persic} \& {Rephaeli}(2022)}]{2022arXiv220802059P}
{Persic}, M., \& {Rephaeli}, Y. 2022, arXiv e-prints, arXiv:2208.02059.
\newblock \doarXiv{2208.02059}

\bibitem[{{Recchia} {et~al.}(2021){Recchia}, {Gabici}, {Aharonian}, \&
  {Niro}}]{Recchia:2021}
{Recchia}, S., {Gabici}, S., {Aharonian}, F.~A., \& {Niro}, V. 2021, \apj, 914,
  135, \dodoi{10.3847/1538-4357/abfda4}

\bibitem[{{Reyes} {et~al.}(2011){Reyes}, {Mandelbaum}, {Gunn}, {Pizagno}, \&
  {Lackner}}]{Reyes:2011}
{Reyes}, R., {Mandelbaum}, R., {Gunn}, J.~E., {Pizagno}, J., \& {Lackner},
  C.~N. 2011, \mnras, 417, 2347, \dodoi{10.1111/j.1365-2966.2011.19415.x}

\bibitem[{{Rojas-Bravo} \& {Araya}(2016)}]{2016MNRAS.463.1068R}
{Rojas-Bravo}, C., \& {Araya}, M. 2016, \mnras, 463, 1068,
  \dodoi{10.1093/mnras/stw2059}

\bibitem[{{Ruszkowski} {et~al.}(2017){Ruszkowski}, {Yang}, \&
  {Zweibel}}]{Ruszkowski:2017}
{Ruszkowski}, M., {Yang}, H.-Y.~K., \& {Zweibel}, E. 2017, \apj, 834, 208,
  \dodoi{10.3847/1538-4357/834/2/208}

\bibitem[{{Salem} {et~al.}(2016){Salem}, {Bryan}, \& {Corlies}}]{Salem:2016}
{Salem}, M., {Bryan}, G.~L., \& {Corlies}, L. 2016, \mnras, 456, 582,
  \dodoi{10.1093/mnras/stv2641}

\bibitem[{{Semenov} {et~al.}(2022){Semenov}, {Kravtsov}, \&
  {Diemer}}]{Semenov:2022}
{Semenov}, V.~A., {Kravtsov}, A.~V., \& {Diemer}, B. 2022, \apjs, 261, 16,
  \dodoi{10.3847/1538-4365/ac69e1}

\bibitem[{{Strong} {et~al.}(2010){Strong}, {Porter}, {Digel},
  {J{\'o}hannesson}, {Martin}, {Moskalenko}, {Murphy}, \&
  {Orlando}}]{strong:2010.milky.way.sub.calorimetric.by.factor.hundred}
{Strong}, A.~W., {Porter}, T.~A., {Digel}, S.~W., {et~al.} 2010, \apjl, 722,
  L58, \dodoi{10.1088/2041-8205/722/1/L58}

\bibitem[{{Su} {et~al.}(2020){Su}, {Hopkins}, {Hayward}, {Faucher-Gigu{\`e}re},
  {Kere{\v{s}}}, {Ma}, {Orr}, {Chan}, \& {Robles}}]{su:turb.crs.quench}
{Su}, K.-Y., {Hopkins}, P.~F., {Hayward}, C.~C., {et~al.} 2020, \mnras, 491,
  1190, \dodoi{10.1093/mnras/stz3011}

\bibitem[{{Ter Haar}(1950)}]{TerHaar:1950}
{Ter Haar}, D. 1950, Reviews of Modern Physics, 22, 119,
  \dodoi{10.1103/RevModPhys.22.119}

\bibitem[{{Tibaldo}(2014)}]{tibaldo.2014:diffuse.gamma.ray.cr.profile.constraints}
{Tibaldo}, L. 2014, Brazilian Journal of Physics, 44, 600,
  \dodoi{10.1007/s13538-014-0221-y}

\bibitem[{{Tibaldo} {et~al.}(2021){Tibaldo}, {Gaggero}, \&
  {Martin}}]{tibaldo.2021:diffuse.gamma.ray.cr.profile.constraints}
{Tibaldo}, L., {Gaggero}, D., \& {Martin}, P. 2021, Universe, 7, 141,
  \dodoi{10.3390/universe7050141}

\bibitem[{{Tibaldo} {et~al.}(2015){Tibaldo}, {Digel}, {Casandjian},
  {Franckowiak}, {Grenier}, {J{\'o}hannesson}, {Marshall}, {Moskalenko},
  {Negro}, {Orlando}, {Porter}, {Reimer}, \&
  {Strong}}]{tibaldo.2015:diffuse.gamma.ray.cr.profile.constraints}
{Tibaldo}, L., {Digel}, S.~W., {Casandjian}, J.~M., {et~al.} 2015, \apj, 807,
  161, \dodoi{10.1088/0004-637X/807/2/161}

\bibitem[{{Trotta} {et~al.}(2011){Trotta}, {J{\'o}hannesson}, {Moskalenko},
  {Porter}, {Ruiz de Austri}, \& {Strong}}]{Trotta:2011}
{Trotta}, R., {J{\'o}hannesson}, G., {Moskalenko}, I.~V., {et~al.} 2011, \apj,
  729, 106, \dodoi{10.1088/0004-637X/729/2/106}

\bibitem[{{Uhlig} {et~al.}(2012){Uhlig}, {Pfrommer}, {Sharma}, {Nath},
  {En{\ss}lin}, \& {Springel}}]{Uhlig:2012}
{Uhlig}, M., {Pfrommer}, C., {Sharma}, M., {et~al.} 2012, \mnras, 423, 2374,
  \dodoi{10.1111/j.1365-2966.2012.21045.x}

\bibitem[{{Werk} {et~al.}(2013){Werk}, {Prochaska}, {Thom}, {Tumlinson},
  {Tripp}, {O'Meara}, \& {Peeples}}]{Werk:2013}
{Werk}, J.~K., {Prochaska}, J.~X., {Thom}, C., {et~al.} 2013, \apjs, 204, 17,
  \dodoi{10.1088/0067-0049/204/2/17}

\bibitem[{{Werk} {et~al.}(2014){Werk}, {Prochaska}, {Tumlinson}, {Peeples},
  {Tripp}, {Fox}, {Lehner}, {Thom}, {O'Meara}, {Ford}, {Bordoloi}, {Katz},
  {Tejos}, {Oppenheimer}, {Dav{\'e}}, \& {Weinberg}}]{Werk:2014}
{Werk}, J.~K., {Prochaska}, J.~X., {Tumlinson}, J., {et~al.} 2014, \apj, 792,
  8, \dodoi{10.1088/0004-637X/792/1/8}

\bibitem[{{Wiener} {et~al.}(2017){Wiener}, {Pfrommer}, \& {Peng
  Oh}}]{Wiener:2017}
{Wiener}, J., {Pfrommer}, C., \& {Peng Oh}, S. 2017, \mnras, 467, 906,
  \dodoi{10.1093/mnras/stx127}

\bibitem[{{Yang} {et~al.}(2016){Yang}, {Aharonian}, \&
  {Evoli}}]{yang.2016:diffuse.gamma.ray.cr.profile.constraints}
{Yang}, R., {Aharonian}, F., \& {Evoli}, C. 2016, \prd, 93, 123007,
  \dodoi{10.1103/PhysRevD.93.123007}

\bibitem[{{Yoast-Hull} {et~al.}(2013){Yoast-Hull}, {Everett}, {Gallagher}, \&
  {Zweibel}}]{yoast.hull:2013.m82.electron.calorimeter.but.not.proton}
{Yoast-Hull}, T.~M., {Everett}, J.~E., {Gallagher}, J.~S., I., \& {Zweibel},
  E.~G. 2013, \apj, 768, 53, \dodoi{10.1088/0004-637X/768/1/53}

\bibitem[{{Yoast-Hull} {et~al.}(2014){Yoast-Hull}, {Gallagher}, {Zweibel}, \&
  {Everett}}]{2014ApJ...780..137Y}
{Yoast-Hull}, T.~M., {Gallagher}, J.~S., I., {Zweibel}, E.~G., \& {Everett},
  J.~E. 2014, \apj, 780, 137, \dodoi{10.1088/0004-637X/780/2/137}

\bibitem[{{Zhang} {et~al.}(2019){Zhang}, {Peng}, \&
  {Wang}}]{2019ApJ...874..173Z}
{Zhang}, Y., {Peng}, F.-K., \& {Wang}, X.-Y. 2019, \apj, 874, 173,
  \dodoi{10.3847/1538-4357/ab0ae2}

\end{thebibliography}
\end{document}